\documentclass{llncs}

\usepackage{graphicx}
	\graphicspath{{./fig/}}
\usepackage{amsmath}
\usepackage{amsfonts}
\usepackage{amssymb}
\usepackage{bm}
\usepackage{mathtools}
\usepackage{nicefrac}
\usepackage{widetable}
\usepackage{booktabs}
\usepackage{minitoc} % toc in each chapter
\usepackage[toc,page]{appendix} % appendix stuff
\usepackage{hyperref} % URLs
\usepackage{soul}
\usepackage{esvect}
\usepackage{pgfgantt}
\usepackage{hyperref}
\usepackage{makeidx}
\usepackage[misc]{ifsym}
\usepackage{tikz} % Graphical models
	\usetikzlibrary{bayesnet} % Graphical models
    \usetikzlibrary{arrows.meta}
\usepackage{multirow}
\usepackage{caption}
\usepackage{gensymb} % \degree symbol

\newcommand{\vt}[1]{{\bf#1}} % vector (text)
\newcommand{\vm}[1]{\bm{#1}} % vector (math)

\begin{document}

%=======================================================================
% TITLE & AUTHORS
%=======================================================================
\title{Groupwise Multimodal Image Registration\\using Joint Total 
Variation}

\author{Mikael Brudfors \and Ya\"{e}l Balbastre \and John Ashburner}
\institute{The Wellcome Centre for Human Neuroimaging, UCL, London,
UK\\ \email{\{mikael.brudfors.15,y.balbastre,j.ashburner\}@ucl.ac.uk}}

\maketitle

%=======================================================================
% ABSTRACT
%=======================================================================
\begin{abstract}

In medical imaging it is common practice to acquire a wide range of
modalities (MRI, CT, PET, etc.), to highlight different structures or
pathologies. As patient movement between scans or scanning session is
unavoidable, registration is often an essential step before any
subsequent image analysis. In this paper, we introduce a cost function
based on joint total variation for such multimodal image registration.
This cost function has the advantage of enabling principled, groupwise
alignment of multiple images, whilst being insensitive to strong
intensity non-uniformities. We evaluate our algorithm on rigidly
aligning both simulated and real 3D brain scans. This validation shows
robustness to strong intensity non-uniformities and low registration
errors for CT/PET to MRI alignment. Our implementation is publicly 
available at 
\url{https://github.com/brudfors/coregistration-njtv}.

\end{abstract}

%=======================================================================
% INTRODUCTION
%=======================================================================
\section{Introduction}

This paper concerns a fundamental task in medical image analysis:
\emph{intermodal} image registration. Its aim is to align scans that
represent the same object, but have been acquired with different
modalities (e.g., multiple repeats of the same MRI sequence, CT, PET).
For \emph{intramodal} registration (e.g., MRIs of the same sequence),
the difference between scans can be assumed to be independent with
Gaussian noise, so that the cost function reduces to the sum of squared
differences \cite{gerlot1992registration,ashburner1997incorporating}. In
contrast, the challenge in intermodal alignment comes from the fact that
the scans are now no longer repeated measures of the same signal,
precluding the use of a simple model of `measurement error'. However,
the complementarity of the information in multimodal data are of crucial
importance in a wide array of applications, from medical diagnosis to
radiotherapy planning. Over the years, many automated registration
algorithms have therefore been developed to tackle the problem of
automated intermodal image alignment
\cite{hill2001medical,oliveira2014medical}.

Automated registration methods often optimise a transformation parameter
of some cost function. The challenge lies in finding a cost function
that has its optimum when images are perfectly aligned. The most
commonly used cost functions are based on intensity cross-correlation
\cite{lewis1995fast,cideciyan1995registration,roche1998correlation},
intensity differences
\cite{hajnal1995detection,woods1998automated,myronenko2010intensity} and
information theory. The most popular functional from information theory
is mutual information (MI;
\cite{collignon1995automated,viola1997alignment,wells1996multi}), which
considers voxels as independent conditioned on a joint intensity
distribution. This distribution is often encoded non-parametrically from
the joint image intensity histogram \cite{collignon1995automated}, but
parametric Gaussian mixture models have also been used
\cite{orchard2009registering}. Normalised mutual information (NMI;
\cite{studholme1999overlap}) was introduced to remove the dependency of
MI to the size of the overlap between field-of-views (FOV). MI-based
cost-functions have been shown to be robust and accurate for medical
image registration \cite{pluim2003mutual}. However, they can fail in the
face of large intensity inconsistencies
\cite{saad2009new,greve2009accurate}, which can be caused, e.g., by
non-homogeneous transmission or reception of the MR signal. To reduce
the dependency on the image intensities, registration approaches based
on aligning edges have been investigated, which include gradient
magnitude correlation \cite{maintz1996comparison}, Canny filters
\cite{orchard2007globally} and normalised gradients dot product
\cite{haber2006intensity,snape2016robust}.

In this paper we propose an edge-based cost function that is groupwise,
finding its optimum when several gradient magnitude images are in
alignment. In contrast to pairwise methods, groupwise registration
defines a cost function over all images to be aligned. Such an
approach should, in principle, lead
to more optimal alignment due to reduced bias and increased number of
registration features
\cite{wachinger2007three,spiclin2012groupwise,polfliet2018intrasubject}.
Our cost function introduces the joint total variation (JTV) functional 
in the
context of image registration. This functional has previously been used 
for
image reconstruction, first in computer vision
\cite{bresson2008fast,wu2010augmented} and then in medical imaging
\cite{huang2014fast,brudfors2018mri}. We evaluate our method on both
simulated and real brain scans. This validation shows robustness to
strong intensity non-uniformities (INUs) and low registration errors for
groupwise, multimodal alignment.

% I think one advantage over ssq over gradients or other edge-based 
% functional, is that it is less sensitive to outliers (like the 
% median compared to the mean). If it was a journal paper, you probably
% would need to compare against such a 'least-squares' version. Maybe 
% a dataset with pathology would be useful to show this robustness.
% (just food for thoughts here)

%=======================================================================
% METHODS
%=======================================================================
\section{Methods}

\noindent \textbf{Total Variation.} The total variation (TV) of a
differentiable function $f:\Omega \subset \mathbb{R}^D \rightarrow
\mathbb{R}$ is the integral of the $\ell_2$-norm of its gradient:
\begin{align}
\textstyle
\small
\operatorname{TV}(f) = \lambda \int_\Omega \left\lVert 
\vm{\nabla} f
(\vt{x}) \right\rVert_2 \mathrm{d}\vt{x},
\end{align}
where $\lambda$ is a scaling parameter that relates to the Laplace
distribution\footnote{A Laplace distribution is given by $p(x) \propto
\exp \left( -\frac{|x|}{b}\right)$, with variance $\sigma^2 = 2b^2$. TV
can be interpreted as a single-sided Laplace distribution over gradient
magnitudes, with $\lambda = \frac{1}{b} = \frac{\sqrt{2}}{\sigma}$.},
and $D=3$ for volumetric medical images. Patient images of different 
modalities can be conceptualised as a multi-channel acquisition. For 
such a vector-valued function,
where the value domain is $\mathbb{R}^C$, two TV functionals that can be
devised are:
\begin{align}
\textstyle
\small
\operatorname{CTV}(\vt{f}) &= \sum_{c=1}^C \lambda_c 
\operatorname{TV}(f_c),\\
\operatorname{JTV}(\vt{f}) &= \int_\Omega \left\lVert 
\left(\lambda_c \vm{\nabla} f_c
(\vt{x})\right)_{1\leqslant c \leqslant 
C} \right\rVert_2 \mathrm{d}
\vt{x}.
\label{eq:jtv}
\end{align}
Here, CTV denotes colour total variation, which considers channels
independently \cite{blomgren1998color}, whereas JTV denotes joint total
variation that applies the norm to the joint gradients over all channels
\cite{sapiro1996anisotropic}. By assuming that each channel is composed
with a rigid transformation $\vt{R}_c:\Omega\rightarrow\Omega$, it can
be shown that CTV is unsuitable as a cost function for image
registration. Defining $\vt{u} = \vt{R}_c\vt{x}$, integration by
substitution gives:
\begin{align}
\textstyle
\small
\int_{\Omega_x} f_c(\vt{x}) \mathrm{d} \vt{x} = \int_{\Omega_u} 
f_c(\vt{u}) 
|\vt{J}| \mathrm{d}
\vt{u},
\end{align}
where $|\vt{J}|$ is the absolute value of the determinant of the
Jacobian matrix of $\vt{u}$. As the determinant of a rigid
transformation is one, we get $|\vt{J}| = 1$. Hence, the value of an
individual TV term does not change with the application of a rigid
transformation, and therefore neither does the CTV term.\\

\begin{figure}[h!]
\centering
\includegraphics[width=\textwidth]{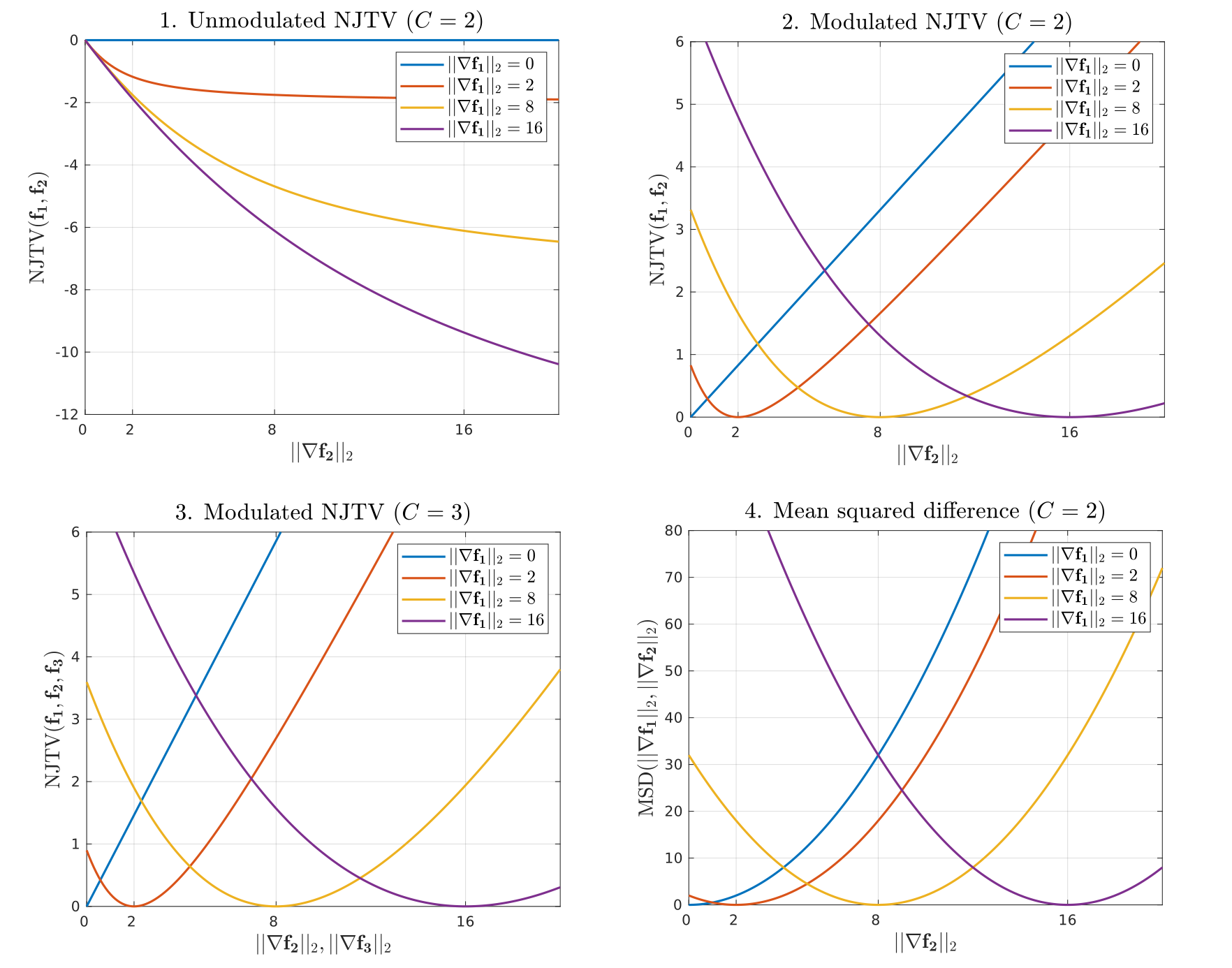}
\caption{Behaviour of the NJTV cost function in \eqref{eq:njtv}.
\textbf{(1)} By computing the NJTV for two voxels (${\bf f}_1$, ${\bf
f}_2$) and changing the value of ${\bf f}_2$, whilst keeping the
gradient magnitude ($|| \nabla {\bf f_1} ||_2$) fixed (for different
values: $0,2,8,16$), it can be seen that NJTV it not minimised for the
correct $|| \nabla {\bf f_1} ||_2$. \textbf{(2)} If we instead modulate
by the square root of $C$, the correct minimum is reached. \textbf{(3)}
Furthermore, computing modulated NJTV in a groupwise setting ($C=3$)
also finds the correct minimum. \textbf{(4)} Mean squared difference
between $|| \nabla {\bf f_1} ||_2$ and $|| \nabla {\bf f_2} ||_2$ is
also minimised, but the curvatures are all the same. For NJTV, the
curvatures vary: for values further from zero, the algorithm has more
flexibility in the gradient magnitudes it matches with (an edge should
match with an edge, but it matters less how strong the matching edges
are); in contrast, if there is no gradient magnitude at a voxel (no
edge), it reverts to a regular $\ell_1$ type penalty.}
\label{fig:modulation}
\end{figure}

\noindent \textbf{Normalised Joint Total Variation.} Images are not
continuous but discrete, and can be defined on non-overlapping domains.
Let $\mathcal{F} = \{{\bf f}_1, \dots, {\bf f}_C\}$ be a set of discrete
images representing the same object, which may have different FOV and
numbers of voxels. These images are made to represent continuous
vector-valued signals by interpolation. An arbitrary image is chosen as
fixed (e.g., $c=1$), and a mapping from the moving images to the fixed
is given by:
\begin{align}
\textstyle
\small
\vm{\xi}_{c}(\vt{x}) = \vt{I}_{3,4} \vt{M}_{1}^{-1} 
\vt{R}_{c} 
\vt{M}_{c} \begin{bmatrix}
\vt{x} \\
1
\end{bmatrix}, \quad c=2,\ldots,C, \quad
\vt{I}_{3,4} = \begin{bmatrix}
1 & 0 & 0 & 0 \\
0 & 1 & 0 & 0 \\
0 & 0 & 1 & 0
\end{bmatrix},
\end{align}
% TODO . I don't think you need to introduce the orientatiaon matrices 
%if the
%   f_c are continuous functions defined by inteprolation. In that 
%%%case, the orientation matrices are 'embedded' in the interpolation 
%   function.
where $\vt{M}_{c}$ is the $c$th image's subject voxel-to-world mapping
(read from the scan's NIfTI header). The JTV of the aligned signal can
then be written as:
\begin{align}
\textstyle
\small
\operatorname{JTV}(\mathcal{F},\mathcal{R}) = 
\int_{\Omega} 
\left\lVert 
\left(\lambda_c \vm{\nabla} f_c
\left( \vm{\xi}_{c}^{-1}(\vt{y}) \right)\right)_{1\leqslant c 
\leqslant 
C} \right\rVert_2 \mathrm{d}
\vt{y},
\end{align}
where $\mathcal{R} = (\vt{R}_{1}, \dots, \vt{R}_{C})$ are the set of
rigid-body transformations to be estimated (except $\vt{R}_{1} = {\bf
I}$) and $\Omega_{f_1}$ is the domain of the fixed image. Values outside
of the fixed image's FOV are nulled, so that the JTV term only involves
observed voxels. As interpolation has a significant impact on the
gradients shape, it is prevented from biasing the optimum by removing
the individual TV terms from the cost function in \eqref{eq:jtv},
arriving at the proposed, normalised joint total variation (NJTV) cost
function\footnote{Here, a parallel can be drawn to the negative mutual
information, which can be computed as the difference of the joint and
individual entropies: $\operatorname{MI}(f_1,f_2) =
-\left(\operatorname{H}[f_1,f_2] - \operatorname{H}[f_1] -
\operatorname{H}[f_2]\right)$.}:
\begin{align}
\textstyle
\small
\operatorname{NJTV}({\mathcal{F},\mathcal{R}}) =  
\sqrt{C} \times \operatorname{JTV}(\mathcal{F},\mathcal{R}) - 
\operatorname{CTV}(\mathcal{F},\mathcal{R}).
\label{eq:njtv}
\end{align}
Note that the JTV term has been modulated with the square root of the
number of channels. This modulation is necessary for the cost function
to find its optimum when all gradient magnitudes are in alignment (more
details in Fig. \ref{fig:modulation}); that is, to be applicable to
image registration.\\

\noindent \textbf{Lie Groups and Rigid-Body Transformations.} We here
consider rigid-body transforms in terms of their membership of the
special Euclidean group in three  dimensions SE(3). This group is a Lie
group and can be equivalently encoded by its Lie algebra
$\mathfrak{se}(3)$ \cite{woods2003characterizing}. Working with the Lie
group representation of SE(3) gives a lower-dimensional, linear
representation for rigid body motion. %\footnote{For example, a rotation
%matrix $\vt{R}$ in the SE(3) group has nine parameters (constrained by
%$\vt{R}\tr \vt{R}=I$ and $\text{det}(\vt{R}) = +1$). However, it has
%only three degrees of freedom; Lie algebra enables working with this
%three parameter representation.}.
An orthonormal basis of the Lie algebra $\mathfrak{se}(3)$ is:
\begin{align*}
\textstyle
\scriptsize
\mathcal{B} =
\begin{Bmatrix}
&\begin{pmatrix}
0 & 0 & 0 & 1 \\
0 & 0 & 0 & 0 \\
0 & 0 & 0 & 0 \\
0 & 0 & 0 & 0
\end{pmatrix},
&\begin{pmatrix}
0 & 0 & 0 & 0 \\
0 & 0 & 0 & 1 \\
0 & 0 & 0 & 0 \\
0 & 0 & 0 & 0
\end{pmatrix},
&\begin{pmatrix}
0 & 0 & 0 & 0 \\
0 & 0 & 0 & 0 \\
0 & 0 & 0 & 1 \\
0 & 0 & 0 & 0
\end{pmatrix},
&\frac{1}{2}\begin{pmatrix}
0 & 1 & 0 & 0 \\
\text{-}1 & 0 & 0 & 0 \\
0 & 0 & 0 & 0 \\
0 & 0 & 0 & 0
\end{pmatrix},
&\frac{1}{2}\begin{pmatrix}
0 & 0 & 1 & 0 \\
0 & 0 & 0 & 0 \\
\text{-}1 & 0 & 0 & 0 \\
0 & 0 & 0 & 0
\end{pmatrix},
&\frac{1}{2}\begin{pmatrix}
0 & 0 & 0 & 0 \\
0 & 0 & 1 & 0 \\
0 & \text{-}1 & 0 & 0 \\
0 & 0 & 0 & 0
\end{pmatrix}
\end{Bmatrix}.
\end{align*}
A 3D rigid-body transform can be encoded by a vector $\vt{q}_c \in
\mathbb{R}^6$ and recovered by matrix exponentiating the Lie algebra
representation:
\begin{align}
\textstyle
\small
\vt{R}_{_c} = \exp\left(\sum_{i=1}^6 q_{ci} \mathcal{B}_i\right).
\end{align}
Conversely, by using a matrix logarithm, the encoding of a rigid-body
matrix can be obtained by projecting it on the algebra:
\begin{align}
\textstyle
\small
q_i = 
\operatorname{Tr}\left(\mathcal{B}_i\log(\vt{R}_{c})^\mathrm{T}\right).
\end{align}\\

\noindent \textbf{Implementation Details.} The NJTV cost function in
\eqref{eq:njtv} is optimised using Powell's method
\cite{press2007numerical}. Powell's method is an algorithm for finding a
local minimum of a function by repeated 1D line-searches, where the
function need not be differentiable, and no derivatives are taken. For
improved runtime we compute the gradient magnitudes in \eqref{eq:njtv}
at the start of the algorithm, and interpolate them using second order
b-splines. To avoid local optima and further improve runtime, a two-step
coarse-to-fine scheme is used. The images are initially downsampled by a
factor of eight and then registered. The algorithm is then run again
using the parameters estimated from the previous registration (a warm
start). The variable voxel size of the input images are accounted for in
the computation of the gradients, by dividing each gradient direction
with its voxel width. The scaling parameters,
$\lambda_1,\ldots,\lambda_C$, which normalise the cost function across
modalities, are estimated from each individual image's intensity
histogram. If an image contains only positive values (e.g., an MR
image), a two-class Rician mixture model is used and the scaling
parameter is set as the mean of the non-background class. If an image
also contains negative values (e.g., a CT image), a Gaussian mixture
model is used instead. In order for the gradient magnitude to be
independent from the data unit, the scaling parameter is set as the
absolute difference between the mean of the background class and the
mean of the foreground class. A random jitter is also introduced to the
sampling grid of the fixed image to reduce interpolation artefacts
\cite{unser2003stochastic}.\\

%=======================================================================
% VALIDATION
%=======================================================================
\begin{figure*}[h!]
\centering
\includegraphics[width=\textwidth]{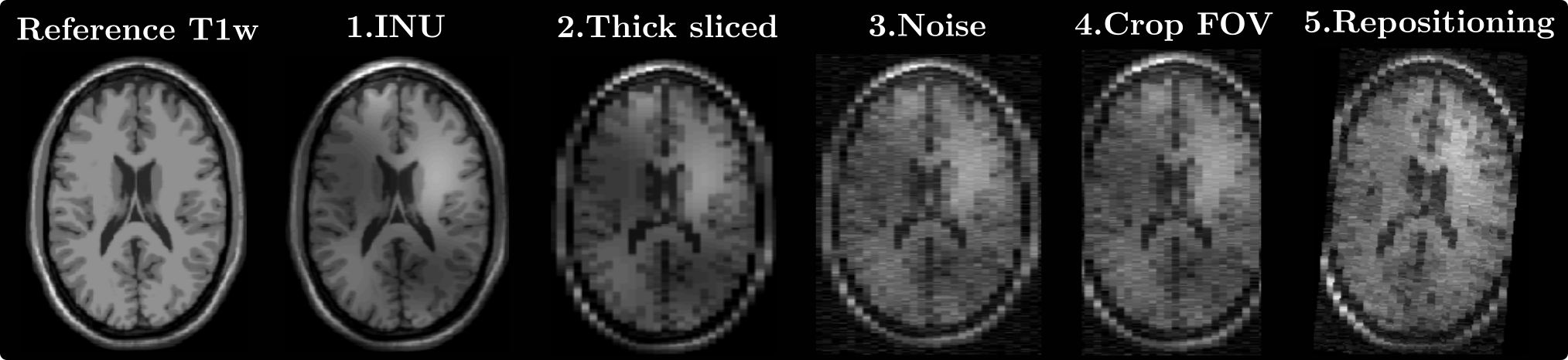}
\caption{The BrainWeb simulation process, where one simulation includes
a T1w, T2w and PDw scan. (1) Intensity non-uniformity (INU) is 
simulated by sampling smooth
multiplicative scalar fields from a multivariate Gaussian distribution.
(2) Thick-sliced data is simulated by downsampling the images, in one
direction, by a factor 1 to 6. (3) A percentage of Rician noise is
added, chosen uniformly between 0\% and 50\% of the maximum image
intensity. (4) Partial brain coverage is simulated by cropping 20 mm on
both sides in a randomly chosen anatomical direction. (5) Rigid
repositioning is applied by uniformly sampled $x$-, $y$-,
$z$-translations (between -50 and 50 mm) and $x$-, $y$-, $z$-rotations
(between -15 and 15 degrees).}
\label{fig:simulation}
\end{figure*}

\section{Validation}

\noindent \textbf{Registering BrainWeb Simulations.} This section
compares NJTV against other common cost functions, using non-degraded 1
mm isotropic T1-weighted (T1w), T2-weighted (T2w) and PD-weighted (PDw)
images from the BrainWeb
simulator\footnote{\url{brainweb.bic.mni.mcgill.ca/brainweb}}
\cite{cocosco1997brainweb}. A series of random degradations are applied
to these reference scans, in order to make them more similar to
clinical-grade data. These degradations are followed by a known rigid
repositioning, which allows for a ground-truth comparison. Figure
\ref{fig:simulation} details this process. The comparison includes the
cost functions implemented in the co-registration routine of
SPM12\footnote{\url{fil.ion.ucl.ac.uk/spm/software/download}}: MI
\cite{wells1996multi,maes1997multimodality}, NMI
\cite{studholme1999overlap}, entropy correlation coefficient (ECC;
\cite{maes1997multimodality}), and normalised cross correlation (NCC;
\cite{lewis1995fast}). For NJTV, the alignment is optimised in a
groupwise setting, whilst for the rest, one image is set as reference
and all other images are aligned with this fixed reference. All cost 
functions use the the same optimisation stopping criteria and 
coarse-to-fine scheme (the defaults in SPM12). Each
transform is encoded by three translations (in mm) and three Euler
angles (in degrees). In total, $2,000$ simulations were performed. For
each simulation, the error was computed between the estimated
transformation parameters and the known ground-truths. The geometric
mean and geometric standard deviation of absolute errors were then
computed for each cost function. To evaluate the impact of different
corruption parameters, a linear model was fitted to the log of the
absolute translation errors generated by each method; with noise level,
downsampling factor, INU magnitude and simulated offset as regressors.
The corresponding maximum-likelihood slopes are written as $\beta_n$,
$\beta_d$, $\beta_i$ and $\beta_o$.

\begin{figure*}[h!]
\centering
\includegraphics[width=\textwidth]{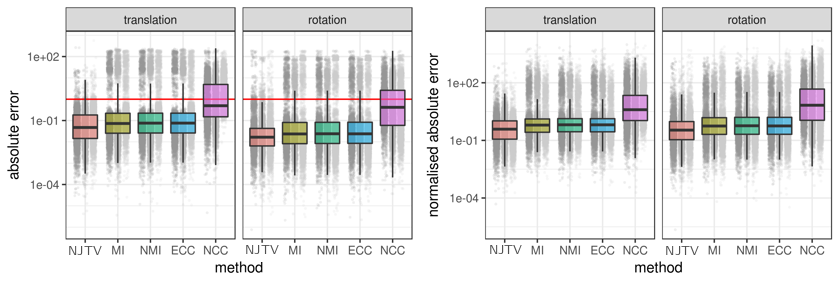}
\caption{Absolute translation (mm) and rotation (deg) errors obtained
from 2,000 BrainWeb simulations. Individual errors along the x, y, and z
directions are plotted in different shades of grey, whereas the boxplot
was computed from the pooled data. The vertical axis is in log-scale, so
that higher points represent a greater error. The plot on the left shows
absolute errors: errors below the red horizontal line are less than 1 mm
or 1 deg. The plot on the right shows errors after normalisation by the
geometric mean across methods.}
\label{fig:errors}
\end{figure*}

\begin{figure*}[h!]
\centering
\includegraphics[width=\textwidth]{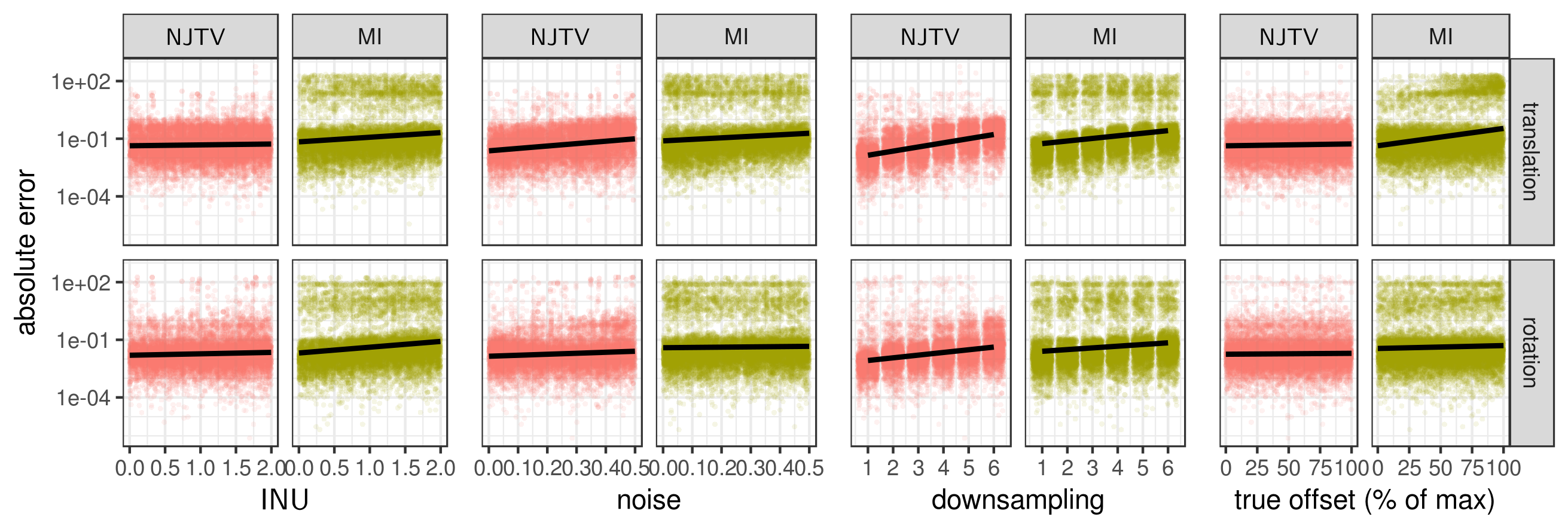}
\caption{Absolute translation (mm) and rotation (deg) errors vs. INU
strength, noise level, downsampling factor and true offset. Offsets are
defined as percentages of the maximum simulated shift (50 mm / 15 deg).
The vertical axis is in log-scale. Errors along each dimension are
considered to be independent, and a regression line is plotted in black.
NJTV, in contrast to MI, is immune to bias field and offset.}
\label{fig:linear_fit}
\end{figure*}

The distribution of absolute errors is shown in Fig. \ref{fig:errors}.
NJTV does consistently better ($\mu_t= 0.048$ mm, $\mu_r =
0.019$\degree), and NCC consistently worse ($\mu_t = 0.69$ mm, $\mu_r =
0.40$\degree), than all other approaches (MI: $\mu_t = 0.12$ mm, $\mu_r
= 0.042$\degree; NMI: $\mu_t = 0.12$ mm, $\mu_r = 0.042$\degree; ECC:
$\mu_t = 0.12$ mm, $\mu_r = 0.042$\degree), which are indistinguishable.
Additionally, there are far fewer outliers with NJTV: a cut-off at 1 mm
gives $97\%$ success for NJTV vs. 85\% for MI, NMI and ECC and 60\% for
NCC. The slopes and intercepts of the log-linear fits are provided in
Table \ref{tab:lm}, and illustrated for NJTV and MI in Fig.
\ref{fig:linear_fit}. NJTV is the method most impacted by noise
($\beta_n=2.91$, compared to MI's $\beta_n=1.80$) and downsampling
($\beta_d=0.50$, compared to MI's $\beta_d=0.31$), but the most robust
to INUs ($\beta_i=0.059$, compared to MI's $\beta_i=0.50$) and to
original misalignment ($\beta_o=0.0044$, compared to MI's
$\beta_o=0.041$).\\

\begin{table*}[t]
\fontsize{7}{7.2}\selectfont
\centering
\caption{Linear model fit to the results of the BrainWeb simulations.
Shown are, for each method, the geometric mean (s.d.) of absolute
translation and rotation errors (columns 2 and 3), as well as the
coefficients of the linear fit \texttt{log(abs(error))
{\raise.17ex\hbox{$\scriptstyle\mathtt{\sim}$}} 1 + INU + noise +
downsampling + offset} applied to translation errors (columns 4 to 7).}
\begin{widetable}{\columnwidth}{ c | c c | c c c c c} \toprule
\textbf{Method} & \textbf{Tr. [mm]} & \textbf{Rot. [deg]} & 
\textbf{1} & \textbf{INU} & \textbf{Noise} & \textbf{DS} & 
\textbf{Offset}
\\\midrule
NJTV & 0.048 (6.20) & 0.019 (6.64) & -5.68 & 0.059 &
2.91 & 0.50 & 0.0044\\
MI  & 0.12  (14.4) & 0.042 (16.0) & -5.18 & 0.50 & 1.80 & 0.31 & 0.041\\
NMI & 0.12  (14.5) & 0.042 (15.8) & -5.22 & 0.53 & 1.76 & 0.33 & 0.040\\
ECC & 0.12  (14.6) & 0.042 (15.5) & -5.19 & 0.45 & 2.12 & 0.33 & 0.041\\
NCC & 0.69   (9.5) & 0.40  (12.6) & -1.47 & 0.21 & 0.13 & 0.14 & 0.015
\\\bottomrule
\end{widetable}
\label{tab:lm}
\end{table*}

\noindent \textbf{Registering CT/PET to MRIs.} The seminal RIRE
multimodal registration challenge \cite{west1996comparison} compared a
wide number of methods at rigidly registering CT/PET to MR scans (T1w,
T2w, PDw). Among these were cost functions, such as normalised mutual
information, that are still considered state-of-the-art over two decades
later. In this section, NJTV is used to register the training patient of
the RIRE
dataset\footnote{\url{insight-journal.org/rire/download_training_data.php}}.
 In the original challenge, the algorithms were compared on scans from 
18 held-out test patients. We here use the training patient's scans, 
with known ground-truth corners, as it is currently not possible to 
submit new methods to the RIRE website (to obtain results on the test 
data). Ideally the testing data should have been used; however, this 
validation still gives an idea how NJTV performs on a multimodal 
registration task. Furthermore, as the algorithm does no learning, no 
such parameter optimisation can be done on the training patient. The 
results of the groupwise alignment is shown in Table \ref{tab:rire}. 
For CT/PET to MRI registration, the average of the median errors for 
all combinations of registrations was 2.0 mm, when all images were 
included in the groupwise registration. This alignment took less than 8 
minutes on a modern workstation. When the groupwise alignment used only 
CT and MRIs, the error was 0.8 mm, whilst when only PET and MRIs were 
used it was 2.0 mm. The errors were computed as in 
\cite{west1996comparison}. The best methods from that paper achieved, 
on the test patients' scans, a CT to MRI error below 2 mm, and a PET to 
MRI error of about 3 mm.\\

\begin{table}[t]
\fontsize{7}{7.2}\selectfont
\centering
\caption{NJTV registration errors for the RIRE training patient. Five
scans (T1w, T2w, PDw, CT, PET) were groupwise aligned, rigidly, using
three different combinations of scans: (top) all, (middle) CT and MRIs,
(bottom) PET and MRIs. The median and maximum error were calculated from
the difference between the eight estimated and ground-truth corner
locations from the CT and PET scans, to the MRIs (in the $x$, $y$, $z$
directions, in mm).}
\begin{widetable}{\columnwidth}{c|ccc|ccc} \toprule
Error& \multicolumn{3}{c}{Median} & \multicolumn{3}{|c}{Max}\\\toprule
From/To& T1w & T2w & PDw & T1w & T2w & PDw \\\toprule
CT  & 2.2 & 1.7& 2.3 & 8.5&  7.2& 9.6\\
PET & 1.9& 1.5& 2.0 & 6.5& 8.6& 11.3\\\midrule
CT  & 1.4 & 0.4& 0.7 & 3.3& 2.3& 3.5\\\midrule
PET & 1.9& 1.7& 2.3 & 9.6& 8.8& 11.1\\\bottomrule
\end{widetable}
\label{tab:rire}
\end{table}

\section{Conclusion}

This paper introduced NJTV as a cost function for image registration.
NJTV provides a principled method for performing accurate groupwise
alignment. We show that NJTV is robust to strong INUs, and fails less
often when faced with large misalignments. Powell's method was here used
to perform the NJTV optimisation. This method has the advantage of not
requiring computing derivatives, but is therefore an inefficient
optimisation scheme. Furthermore, it only works for cost functions with
a small number of transformation parameters, such as affine
registration. Future work will therefore investigate more efficient,
derivative-based optimisation techniques, which could allow for
groupwise nonlinear alignment using NJTV.

\subsubsection*{Acknowledgements:} MB was funded by the EPSRC-funded
UCL Centre for Doctoral Training in Medical Imaging (EP/L016478/1) and
the Department of Health’s NIHR-funded Biomedical Research Centre at
University College London Hospitals. YB was funded
by the MRC and Spinal Research Charity through the ERA-NET Neuron joint
call (MR/R000050/1). MB and JA were funded by the EU
Human Brain Project's Grant Agreement No 785907 (SGA2).

%=======================================================================
% REFERENCES
%=======================================================================
\bibliography{bibliography}
\bibliographystyle{ieeetr}

\end{document}